\documentclass[sigconf]{acmart}
\usepackage{multirow}      
\usepackage{multicol} 
\usepackage{booktabs}
\usepackage{algorithm}
\usepackage{algorithmic}
\usepackage{balance} 
\usepackage{url}
\AtBeginDocument{%
  \providecommand\BibTeX{{%
    \normalfont B\kern-0.5em{\scshape i\kern-0.25em b}\kern-0.8em\TeX}}}

\copyrightyear{2021}
\acmYear{2021}
\setcopyright{acmlicensed}\acmConference[MM '21]{Proceedings of the 29th ACM International Conference on Multimedia}{October 20--24, 2021}{Virtual Event, China}
\acmBooktitle{Proceedings of the 29th ACM International Conference on Multimedia (MM '21), October 20--24, 2021, Virtual Event, China}
\acmPrice{15.00}
\acmDOI{10.1145/3474085.3475554}
\acmISBN{978-1-4503-8651-7/21/10}

\begin{document}

\title{Enhancing Knowledge Tracing via Adversarial Training}

%\author{Xiaopeng Guo, Zhijie Huang, Jie Gao, Mingyu Shang, Maojing Shu, Jun Sun}
%\email{xiaopeng.guo@stu.pku.edu.cn,  {zhijiehuang, gaojie2018, mingyu.shang, shumaojing, sunjun}@pku.edu.cn}
%\orcid{0000-0003-1111-2035}
%\affiliation{%
%  \institution{Wangxuan Institute of Computer Technology, Peking University}
%  \streetaddress{No. 128 Zhongguancun North Street, Haidian District, Beijing, 100871, P. R. China}
%  \city{Beijing}
%  \state{}
%  \country{China}
%  \postcode{100871}
%}

\author{Xiaopeng Guo}
\email{xiaopeng.guo@stu.pku.edu.cn}
\orcid{0000-0003-1111-2035}
\affiliation{%
	\institution{Wangxuan Institute of Computer Technology, Peking University}
	\streetaddress{No. 128 Zhongguancun North Street, Haidian District, Beijing, 100871, P. R. China}
	\city{Beijing}
	\state{}
	\country{China}
	\postcode{100871}
}

\author{Zhijie Huang}
\email{zhijiehuang@pku.edu.cn}
\affiliation{%
	\institution{Wangxuan Institute of Computer Technology, Peking University}
	\streetaddress{No. 128 Zhongguancun North Street, Haidian District, Beijing, 100871, P. R. China}
	\city{Beijing}
	\state{}
	\country{China}
	\postcode{100871}
}

\author{Jie Gao}
\email{gaojie2018@pku.edu.cn}
\affiliation{%
	\institution{Wangxuan Institute of Computer Technology, Peking University}
	\streetaddress{No. 128 Zhongguancun North Street, Haidian District, Beijing, 100871, P. R. China}
	\city{Beijing}
	\state{}
	\country{China}
	\postcode{100871}
}

\author{Mingyu Shang}
\email{mingyu.shang@pku.edu.cn}
\affiliation{%
	\institution{Wangxuan Institute of Computer Technology, Peking University}
	\streetaddress{No. 128 Zhongguancun North Street, Haidian District, Beijing, 100871, P. R. China}
	\city{Beijing}
	\state{}
	\country{China}
	\postcode{100871}
}

\author{Maojing Shu}
\email{shumaojing@pku.edu.cn}
\affiliation{%
	\institution{Wangxuan Institute of Computer Technology, Peking University}
	\streetaddress{No. 128 Zhongguancun North Street, Haidian District, Beijing, 100871, P. R. China}
	\city{Beijing}
	\state{}
	\country{China}
	\postcode{100871}
}

\author{Jun Sun}
\authornote{Corresponding Author}
\email{sunjun@pku.edu.cn}
\affiliation{%
	\institution{Wangxuan Institute of Computer Technology, Peking University}
	\streetaddress{No. 128 Zhongguancun North Street, Haidian District, Beijing, 100871, P. R. China}
	\city{Beijing}
	\state{}
	\country{China}
	\postcode{100871}
}

\renewcommand{\shortauthors}{Trovato and Tobin, et al.}

\begin{abstract}
We study the problem of knowledge tracing (KT) where the goal is to trace the students' knowledge mastery over time so as to make predictions on their future performance. Owing to the good representation capacity of deep neural networks (DNNs), recent advances on KT have increasingly concentrated on exploring DNNs to improve the performance of KT. However, we empirically reveal that the DNNs based KT models may run the risk of overfitting, especially on small datasets, leading to limited generalization. In this paper, by leveraging the current advances in adversarial training (AT), we propose an efficient AT based KT method (ATKT) to enhance KT model's generalization and thus push the limit of KT. Specifically, we first construct adversarial perturbations and add them on the original interaction embeddings as adversarial examples. The original and adversarial examples are further used to jointly train the KT model, forcing it is not only to be robust to the adversarial examples, but also to enhance the generalization over the original ones. To better implement AT, we then present an efficient attentive-LSTM model as KT backbone, where the key is a proposed knowledge hidden state attention module that adaptively aggregates information from previous knowledge hidden states while simultaneously highlighting the importance of current knowledge hidden state to make a more accurate prediction. Extensive experiments on four public benchmark datasets demonstrate that our ATKT achieves new state-of-the-art performance. Code is available at: \color{blue} {\url{https://github.com/xiaopengguo/ATKT}}.
\end{abstract}





\maketitle

\section{Introduction}

Thanks to the rapid progress of digital multimedia technologies, online learning platforms such as massive open online courses (MOOCs)\footnote{https://www.mooc.org} and Coursera\footnote{https://www.coursera.org} have received increasingly attention from the public \cite{anderson2014engaging, wang2020fine}. A huge amount of datasets about the learning process of students have been collected through these platforms to facilitate the advance of intelligent education field. Among the topics in this field, knowledge tracing (KT) is considered as an essential task with the goal of using students' historical learning data to model their knowledge mastery over time so as to make predictions on their future performance \cite{zhang2017dynamic, wang2019deep}. Take Fig. 1 as an example. A certain student \emph{s} has sequentially answered four exercises ($e_{1}$-$e_{4}$) where $e_{1}$, $e_{2}$, and $e_{4}$ are correct while $e_{3}$ is wrong, indicating that \emph{s} may have a good mastery of the knowledge skill (KS) ``Square Root'', ``Absolute Value'' and ``Algebraic Simplification'' but is not familiar with the ``Linear Equations'' KS. With the current mastery of each KS, how will the student behave in the following fifth exercise $e_{5}$ with the KS of ``Algebraic Solving''? KT is such an effective technique to address this question. Once the knowledge mastery is acquired through KT, students can find out their weakness in specific KS in time and then carry out targeted exercising \cite{liu2019ekt}; also, it helps online learning platforms to customize better personalized exercise recommendation service for different customers \cite{liu2019ekt, wu2020exercise}.

\begin{figure}[t]
	\centering
	\includegraphics[width=\linewidth]{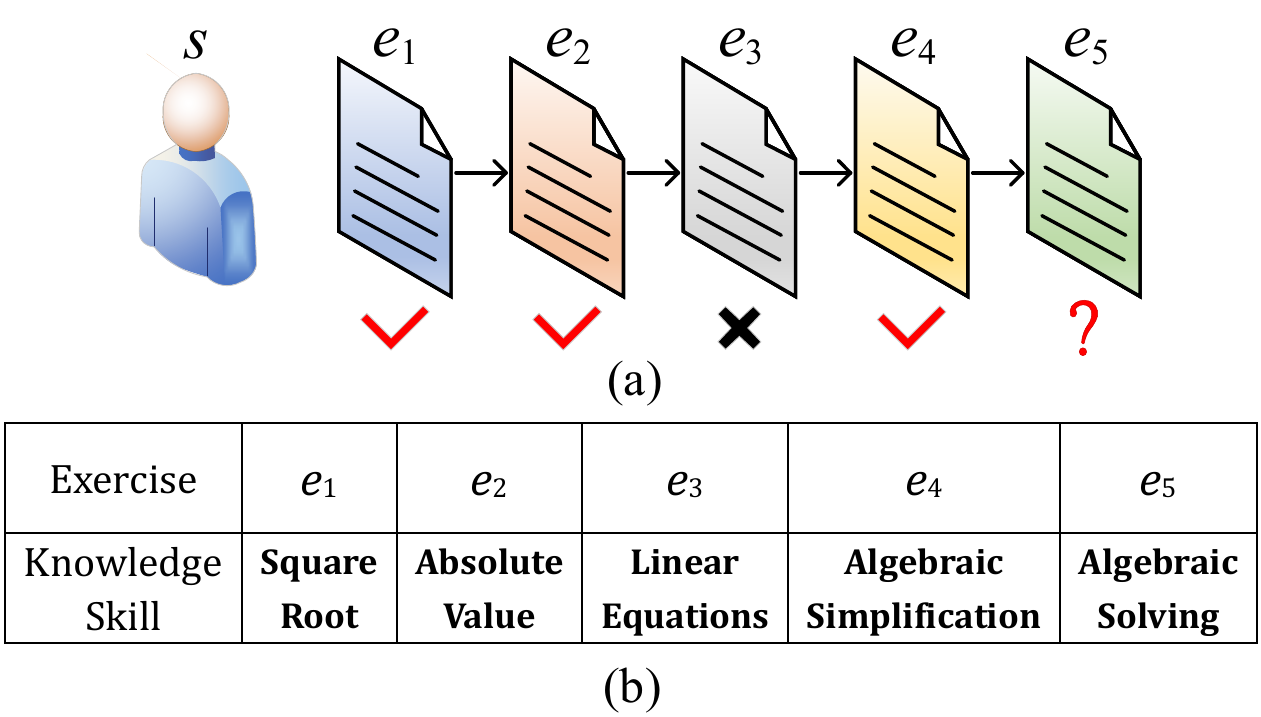}
	\caption{An illustration of exercising process. (a): the exercising process of a certain student. (b): the corresponding KS of each exercise.}
	%	\Description{A woman and a girl in white dresses sit in an open car.}
\end{figure}

Over the few decades, many inspiring attempts have been made to push the improvement of KT, among which the early ones mainly along the research line of probabilistic modelling manner \cite{yeung2018addressing}. Representative examples include \cite{corbett1994knowledge, yudelson2013individualized, hawkins2014learning, kaser2017dynamic, cen2006learning, pavlik2009performance}. While these approaches achieve encouraging progress, they always rely on simplified assumptions such as no forgetting during the learning process  \cite{corbett1994knowledge}, limiting their applications in real scenarios. Moreover, these methods are always hand-crafted engineerings; however, incorporating all potential factors into the KT model is essentially a non-trival task. Owing to the good representation capacity of deep neural networks (DNNs) \cite{lecun2015deep}, some beneficial explorations have been made on KT in recent years \cite{NIPS2015_bac9162b, yeung2018addressing, zhang2017dynamic, c4e83c715e3a43c596941347a55c91d7, ghosh2020context, pandey2020rkt, minn2018deep, wang2019deep, lee2019knowledge}. Roughly speaking, those methods mainly leverage the recurrent neural network (RNN) \cite{hochreiter1997long} or the attention mechanism \cite{vaswani2017attention} or a combination of both to model KT. 

Despite impressive performance have been achieved by these DNNs based methods above, they still may run the risk of overfitting, especially on small-scale datasets. Experimentally, we train a DNNs based KT model \cite{NIPS2015_bac9162b} on a small dataset Statics2011\footnote{https://pslcdatashop.web.cmu.edu/DatasetInfo?datasetId=507} and show the training and validation loss curves in Fig. 2. It can be intuitively observed that with the continuous reduction of the training loss, the validation loss would instead increase after 20 epochs, indicating that the KT model might be overfitted on training data set; thus, the generalization of the KT model on the validation set might be somewhat limited. To push the limit of KT, the generalization of the KT model should be further enhanced. We argue that this can be achieved by leveraging the current advances in \emph{adversarial training} (AT). In fact, as stated in \cite{biggio2013evasion, 42503, 43405}, many DNNs are often vulnerable to \emph{adversarial examples} (AEs), which are generated by adding maliciously designed perturbations on original clean inputs, degrading the performance of DNNs severely. Fortunately, AT provides an efficient way to not only improve the robustness of the DNNs against perturbations but also enhance its generalization over original inputs by training DNNs to correctly classify both of the original inputs and AEs \cite{42503, 45839}. So far, AT has been successfully applied to various fields, ranging from image \cite{43405, tang2019adversarial} to text \cite{45839, liu2020robust}. Our work also draws inspiration from AT. 
\begin{figure}[t]
	\centering
	\includegraphics[width=\linewidth]{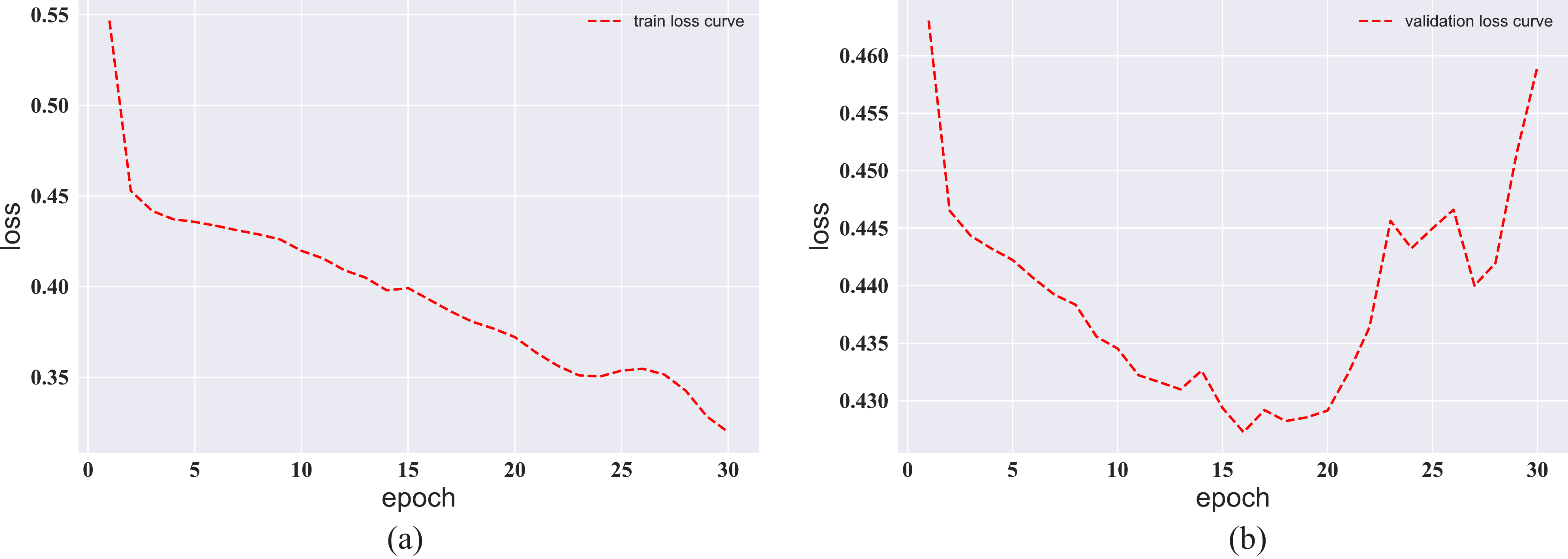}
	\caption{Training and validation loss curves on the small Statics2011 dataset by training a DNNs based KT method $[27]$. (a): the training loss curve and (b): the validation loss curve.}
\end{figure}

In this paper, we contribute from a more general view and explore to leverage the advanced AT to enhance the generalization of KT model for better performance. To this end, we should first determine the input representation of KT. Different from image data that can be represented as continuous pixel values, the raw data in KT is discrete student interactions in the form of (knowledge skill (KS), response) tuples. Both KS and response are similar to words in the text data. In \cite{45839}, the authors use embeddings to represent words, and add perturbations on them to generate AEs to fulfill AT. Motivated by this work, we also employ the embedding to represent student interactions in KT. Specifically, we construct two types of embeddings, i.e., KS embedding and response  embedding to represent the KS and response, respectively. These two embeddings are further concatenated to obtain a full representation of the student interaction as the input representation of KT. 

We then customize an efficient attentive-LSTM model as KT backbone to better implement AT, where the key technical innovation is the proposed knowledge hidden state (KHS) attention module, which aims to adaptively aggregate  information from previous KHSs while simultaneously highlighting the importance of current KHS to make the final prediction. Our design spirit is similar to \cite{liu2019ekt}, which also combines historical information with an attention mechanism to make the future prediction. However, there are two-fold distinct differences between our KHS attention module and the one in \cite{liu2019ekt}. First, instead of using \emph{Cosine Similarities} to measure the importance of all the  historical interactions for the future interaction, we only employ a softmax function to implicitly calculate the attention score of historical interactions. Second, different from directly summing the historical state information in \cite{liu2019ekt}, we not only aggregate information from previous KHSs but also emphasize current KHS. We then concatenate the aggregated KHSs with the current KHS as a composite representation to make a more accurate prediction. With the proposed attentive-LSTM backbone, we add perturbations on the concatenated interaction embeddings to produce AEs as adversaries, and further combine the original inputs to train the model, resulting in distinct enhancement of the generalization of the KT model. We conduct extensive experiments on four public benchmark datasets to evaluate our AT based KT (ATKT) method. Results show that our ATKT achieves new state-of-the-art performance, demonstrating the effectiveness of AT in KT modelling.

To sum up, the main contributations of this work are two-fold:
\begin{itemize}
	\item We highlight the overfitting risks faced by current DNNs based KT models and empirically reveal their relatively limited generalization, especially on small datasets.
	\item We propose ATKT to enhance the generalization of KT model by jointly training the original clean inputs and the corresponding perturbed AEs. To better implement AT, we introduce an efficient attentive-LSTM model as the KT backbone. The key of the proposed KT backbone is a KHS attention module, which can adaptively aggregate all information from previous KHSs while highlighting the importance of current KHS to achieve a more accurate prediction. 
\end{itemize}

\begin{figure*}[t]
	\centering
	\includegraphics[width=\linewidth]{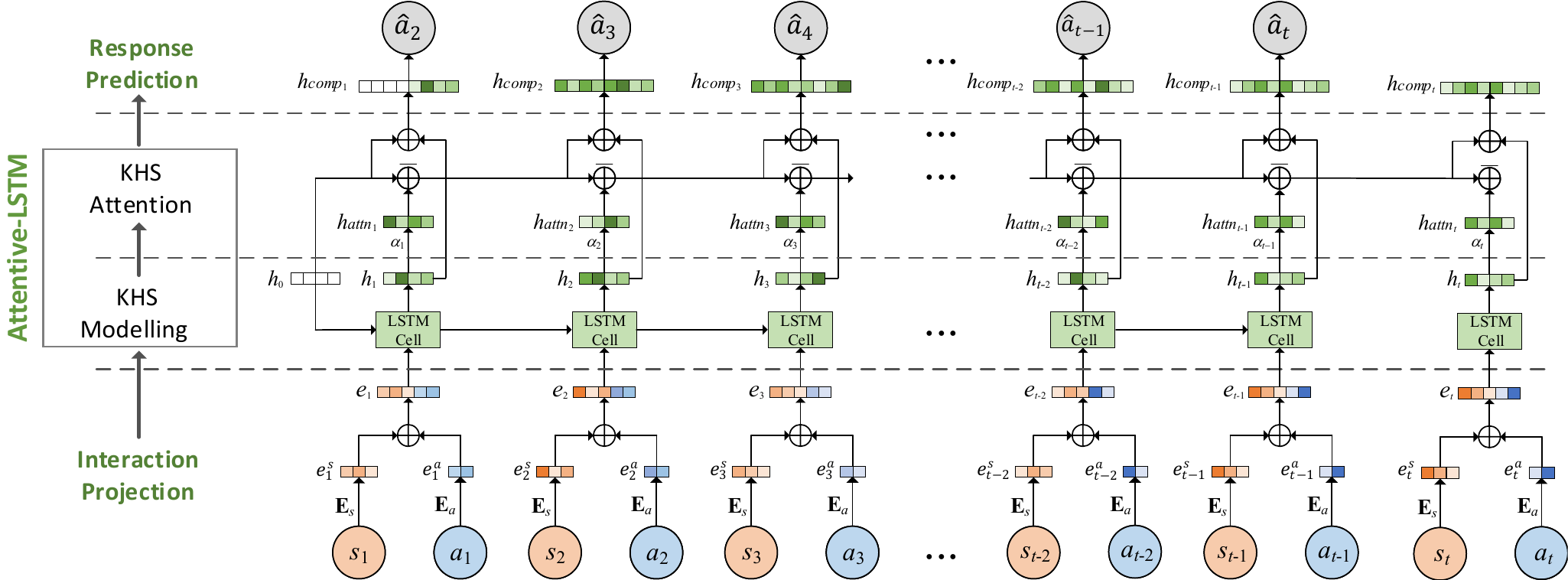}
	\caption{An overview of our entire KT framework, which consists of three components: \emph{Interaction Projection}, \emph{Attentive-LSTM} and \emph{Response Prediction}. \emph{Interaction Projection} is responsible for projecting the student interaction (${s}_{t-1}$, ${a}_{t-1}$) onto the interaction embedding $\boldsymbol{e}_{{t-1}}$; \emph{Attentive-LSTM} is presented as the KT backbone, where a KHS modelling module is first constructed to obtain an initial KHS $\boldsymbol{h}_{{t-1}}$ and then a KHS attention module is carefully designed to adaptively aggregate information from previous KHSs while simultaneously highlighting the importance of current KHS to produce the composite representation $\boldsymbol{h}_{comp_{j}}$ for a more accurate prediction. Finally, $\boldsymbol{h}_{comp_{t-1}}$ is fed into \emph{Response Prediction} to make the final prediction $\hat{a}_{t}$. ${\oplus}$ and $\bar{\oplus}$ denote the concatenation and addition operation, respectively.}
\end{figure*}

\section{Related Work}
\subsection{Knowledge Tracing (KT)}
Previous KT algorithms can be broadly divided into two categories: traditional machine learning methods and DNNs ones. In the first category, Bayesian knowledge tracing (BKT) \cite{corbett1994knowledge} is one of the most representative methods. It assumes the student's knowledge state as a set of binary variables, and then employs a Hidden Markov Model to update the knowledge state based on the student performance on exercises. Several attempts are further made to extend BKT by incorporating other factors, such as slip and guess estimation \cite{d2008more}, exercise difficulty \cite{pardos2011kt} and student personalization \cite{yudelson2013individualized}, et al. With the huge breakthrough of DNNs in image classification \cite{krizhevsky2012imagenet}, the powerful representation capability of DNNs has shifted researchers' attention from the conventional machine learning view to that of DNNs to model KT. Piech et al. \cite{NIPS2015_bac9162b} first introduce DNNs into KT and present a deep KT (DKT) method to employ a RNN to model the student's latent knowledge state through sequentially updated hidden variables, achieving better performance compared with traditional methods. Since then, several approaches have been made to further extend DKT. To name a few, Yeung et al. \cite{yeung2018addressing} point out the problems of input reconstruction failure and prediction inconsistency in DKT, and solve them by prediction-consistent regularization technique. Minn et al. \cite{minn2018deep} bring the information of students' learning ability into DKT. Nagatani et al. \cite{nagatani2019augmenting} take forgetting behavior into account to augment DKT. Zhang et al. \cite{zhang2017incorporating} incorporate more rich features, such as students response time, attempt number and first action to jointly model KT. More recently, attention mechanism \cite{vaswani2017attention} has become popular in KT modelling with the aim of capturing long-term dependencies between interactions \cite{zhang2017dynamic, c4e83c715e3a43c596941347a55c91d7, pandey2020rkt, ghosh2020context}. Unlike these efforts that always incorporate multiple learning factors to improve KT, we focus primarily on enhancing the generalization of the KT model itself, and leverage the advanced AT to achieve this goal.

\subsection{Adversarial Training (AT)}
AT is an efficient regularization technique, which is originally applied to image classification with continuous input data \cite{43405}. Later, realizing that the common one-hot representation of text does not admit perturbations, Miyato et al. \cite{45839} instead define the adversarial perturbations on continuous word embeddings to successfully extend AT to text classification. Dong et al. \cite{dong2020leveraging} further propose a self-learning based method to employ AT to cross-lingual text classification. In addition, some methods have also been proposed to explore the use of AT to model other tasks. For instance, Wu et al. \cite{wu2017adversarial} employ AT in relation extraction within the multi-instance multi-label learning framework. Yang et al. \cite{yang2019improving} study the effects of AT to machine reading comprehension in a multi-perspective analysis manner. Yasunaga et al. \cite{yasunaga-etal-2018-robust} use AT to present and then analyze a neural part-of-speech tagging model. Sato et al. \cite{sato2019effective} discuss the practical usage and benefit of AT in neural machine translation modelling. Though AT reveals promising results on these tasks above, its specific impact on KT has not been thoroughly investigated, which will be the main focus of this paper.

\section{Approach}
We detail our ATKT in this section. We first give a formal definition of KT and then present our attentive-LSTM based KT framework. Finally, we introduce how to implement AT to this framework to improve its generalization so as to push the limit of KT.
\subsection{Problem Formulation}
Let denote a certain student's historical interactions up to time $t-1$ on a specific learning course as $\mathcal{X}=\left\{ \left(\mathrm{s}_{i}, a_{i}\right) \mid i=1, \ldots, t-1 \right\}$, where $s_{i} \in \mathcal{S}$ is the KS index of an exercise that the student attempts at time $i$, and $\mathcal{S}$ denotes the set of KSs with the total number $|\mathcal{S}|$. $a_{i} \in \{0,1\}$ represents the correctness of the response, where $a_{i}=1$ indicates a right response while $a_{i}=0$ is a wrong one. With this definition, KT can be formulated as a binary sequence prediction problem: given $\mathcal{X}$ and the next exercise with KS index $s_{t}$, the goal of KT is to predict the probability of the student answering the exercise correctly, i.e., $P\left(a_{t}=1 \mid \mathcal{X}, s_{t} \right)$.

\subsection{KT framework}
As shown in Fig. 3, our entire KT framework consists of three components, including \textbf{\emph{Interaction Projection}}, \textbf{\emph{Attentive-LSTM}} and \textbf{\emph{Response Prediction}}.

\subsubsection{\textbf{Interaction Projection}}
We use embeddings to project the original KS and response indexes onto the embedding space. Let denote the KS and response embedding matrices as $\mathbf{E}_{s} \in \mathbb{R}^{|\mathcal{S}| \times d_{s}}$ and $\mathbf{E}_{a} \in \mathbb{R}^{2 \times d_{a}}$ respectively, where $d_{s}$ and $d_{a}$ represent the corresponding embedding dimension. Each row in $\mathbf{E}_{s}$ stands for a specific KS while the two rows in $\mathbf{E}_{a}$ represent incorrect and correct answers respectively. Given a certain student interaction with the KS ${s}_{j}$ and response ${a}_{j}$, we can obtain their corresponding embeddings $\emph{\textbf{e}}^{s}_{{j}} \in \mathbb{R}^{1 \times d_{s}}$ and $\emph{\textbf{e}}^{a}_{{j}}\in \mathbb{R}^{1 \times d_{a}}$ through looking up $\mathbf{E}_{s}$ and $\mathbf{E}_{a}$. To acquire a full representation of student's interaction, a natural strategy is to concatenate $\emph{\textbf{e}}^{s}_{j}$ and $\emph{\textbf{e}}^{a}_{j}$ straightforwardly. However, to emphasize the difference between correct and incorrect response, we here adopt an response aware scheme to produce $\emph{\textbf{e}}_{{j}}$ as
\begin{equation}
\emph{\textbf{e}}_{{j}}=\left\{\begin{array}{ll}{\emph{\textbf{e}}^{s}_{j}\oplus \emph{\textbf{e}}^{a}_{j}} & \text { if } \quad a_{j}=1, \\ {\emph{\textbf{e}}^{a}_{j} \oplus \emph{\textbf{e}}^{s}_{j}} & \text { if } \quad a_{j}=0,\end{array}\right.
\end{equation}
where $\oplus$ stands for the concatenation operation and $\emph{\textbf{e}}_{{j}} \in \mathbb{R}^{1 \times (d_{s}+d_{a})}$ is the full representation of student's interaction.

\subsubsection{\textbf{Attentive-LSTM}}
The proposed attentive-LSTM model is used as the KT backbone in this work, which consists of \emph{KHS Modelling } and \emph{KHS Attention} modules.

\emph{KHS Modelling}: With the full representation $\left\{\emph{\textbf{e}}_{1}, \emph{\textbf{e}}_{{2}}, \emph{\textbf{e}}_{{3}}, \ldots, \emph{\textbf{e}}_{{t}} \right\}$ of student's interaction defined in section 3.2.1 above, we intend to capture the  KHS $\emph{\textbf{h}}_{t} \in \mathbb{R}^{d_{h} \times 1}$ at each time step $t$, where $d_{h}$ is the dimension of $\emph{\textbf{h}}_{t}$. Following many previous KT methods \cite{NIPS2015_bac9162b, yeung2018addressing, minn2018deep, zhang2017incorporating, nagatani2019augmenting}, we also adopt the currently popular RNN sequence modelling method to achieve this goal. Specifically, we employ a long short-term memory (LSTM) model \cite{hochreiter1997long} to recurrently update $\emph{\textbf{h}}_{t}$ by receiving the current $\emph{\textbf{e}}_{{t}}$ and the previous $\emph{\textbf{h}}_{t-1}$ as input, i.e.,
\begin{equation}
\emph{\textbf{h}}_{t}\longleftarrow LSTM(\emph{\textbf{e}}_{{t}}, \emph{\textbf{h}}_{t-1}).
\end{equation}
Note that at the first time step $t=1$, $\emph{\textbf{h}}_{t-1}$ is equal to $\emph{\textbf{h}}_{0}$. We initialize it to a zero vector as the initial KHS.

\emph{KHS Attention}: Using the updated $\emph{\textbf{h}}_{t-1}$, one can directly predict the probability that the student correctly answering the next exercise with the KS $s_{t}$. However, the student's interactions at each time step may have different effects on the update of $\emph{\textbf{h}}_{t-1}$. For instance, the KS ``Absolute Value'' in Fig. 1 may have a larger influence since it's the foundation for mastering other more complex KSs, such as ``Algebraic Solving''. Based on this observation, we employ an attention mechanism to $\left\{\boldsymbol{h}_{1}, \boldsymbol{h}_{2}, \ldots, \boldsymbol{h}_{t}\right\}$ to adaptively obtain the importance coefficient of each interaction. Formally, 
\begin{equation}
\boldsymbol{u}_{i}=\tanh \left(\boldsymbol{W}_{w} \boldsymbol{h}_{i}+\boldsymbol{b}_{w}\right),
\end{equation}
\begin{equation}
\alpha_{t}=\frac{\exp \left(\boldsymbol{u}_{i}^{\top} \boldsymbol{u}_{w}\right)}{\sum_{t} \exp \left(\boldsymbol{u}_{i}^{\top} \boldsymbol{u}_{w}\right)}.
\end{equation}
More specifically, ${\boldsymbol{h}}_{i}$ is first fed into a single multi-layer perceptron (MLP) with the learnable weight $\boldsymbol{W}_{w} \in \mathbb{R}^{d_{w} \times d_{h}}$ and bias $\boldsymbol{b}_{w} \in \mathbb{R}^{d_{w} \times 1}$, followed by a tanh activation function to produce the hidden representation ${\boldsymbol{u}}_{i} \in \mathbb{R}^{d_{w} \times 1}$; then, by measuring the similarity between ${\boldsymbol{u}}_{i}$ and ${\boldsymbol{u}}_{w} \in \mathbb{R}^{d_{w} \times 1}$ through the dot product operation, followed by a softmax function, the importance coefficient $\alpha_{t}$ can be generated. Since the previous KHS sequences $\left\{\boldsymbol{h}_{1}, \boldsymbol{h}_{2}, \ldots, \boldsymbol{h}_{t-2}\right\}$ would affect the current  $\boldsymbol{h}_{t-1}$ for predictions differently, we define their total influence  as weighted sum aggregation of all the $\left\{\boldsymbol{h}_{1}, \boldsymbol{h}_{2}, \ldots, \boldsymbol{h}_{t-2}\right\}$, namely, 
\begin{equation}
\hat{\boldsymbol{h}}_{t-2}=\sum_{j=1}^{t-2}  \boldsymbol{h}_{{attn}_{j}},
\end{equation}
\begin{equation}
\boldsymbol{h}_{{attn}_{j}}=\alpha_{j} \boldsymbol{h}_{j}.
\end{equation}
Moreover, we should also highlight the importance of $\boldsymbol{h}_{t-1}$ since it implicitly represents the student's current knowledge mastery and is closely related to the prediction of correctness in the next exercise. Therefore, we concatenate $\hat{\boldsymbol{h}}_{t-2}$ and $\boldsymbol{h}_{t-1}$ as a composite represention $\boldsymbol{h}_{{comp}_{t-1}}$ to make the final prediction,
\begin{equation}
\boldsymbol{h}_{{comp}_{t-1}}=\hat{\boldsymbol{h}}_{t-2} \oplus \boldsymbol{h}_{t-1}.
\end{equation}

\subsubsection{\textbf{Response Prediction}}
This component is to predict the student's response to the next exercise with the KS $s_{t}$ by using the composite represention $h_{{comp}_{t-1}}$. $h_{{comp}_{t-1}}$ goes through a MLP followed by a Sgimoid function to produce the predicted probability $\hat{a}_{t}\in[0,1]$. Binary cross entropy is used to compute the loss between $\hat{a}_{t}$ and the response $a_{t}$, i.e., 
\begin{equation}
l(\hat{a}_{t}, a_{t})=-\left(a_{t} \log \hat{a}_{t}+\left(1-a_{t}\right) \log \left(1-\hat{a}_{t}\right)\right).
\end{equation}

\subsection{ATKT}
We use an attentive-LSTM based framework $\mathcal{F}(; \boldsymbol{\Theta})$ ($\boldsymbol{\Theta}$ is parameter set) introduced above to model KT. Specifically, given a certain student's interaction sequence $\mathcal{X}=\left\{ \left(\mathrm{s}_{i}, a_{i}\right) \mid i=1, \ldots, t \right\}$, we can obtain the interaction embedding set $\mathcal{E}=\left\{\boldsymbol{e}_{i} \mid i=1, \ldots, t \right\}$ using the \emph{Interaction Projection} component. Then, we use $\mathcal{E}$ and $\mathcal{A}=\left\{ {a}_{i} \mid i=2, \ldots, t \right\}$ to update $\boldsymbol{\Theta}$ by minimizing the following loss
\begin{equation}
\boldsymbol{\Theta}\gets \mathcal{L}(\mathcal{E}, \mathcal{A}, \boldsymbol{\Theta}),
\end{equation}
\begin{equation}
\mathcal{L}(\mathcal{E}, \mathcal{A},\boldsymbol{\Theta})=\frac{1}{\left|\mathcal{E}\right|-1} \sum_{t}l\left(\mathcal{F}\left(\mathbf{e}_{t-1}; \boldsymbol{\Theta}\right), a_{t}\right).
\end{equation}
The well-trained model $\mathcal{F}(;\hat{\boldsymbol{\Theta}})$ could be produced by training $\mathcal{F}(; \boldsymbol{\Theta})$ on the whole training dataset. 
Though $\mathcal{F}(;\hat{\boldsymbol{\Theta}})$ could gain satisfactory performance, we believe that its generalization is still limited and can be further enhanced by resorting to the advanced AT. As an efficient regularization technique, AT not only improves the robustness of the model against perturbations but also enhances its generalization over clean inputs. To implement AT successfully, we should first construct AEs. Following \cite{yang2019improving}, we add perturbations on the interaction embedding to achieve this goal. Let ${\boldsymbol{r}}_{i}^{\prime} \in \mathbb{R}^{1 \times (d_{s}+d_{a})}$ be an adversarial perturbation for the $i$-th interaction embedding $\boldsymbol{e}_{i}$ in $\mathcal{E}$, then the perturbed interaction embedding $\boldsymbol{e}_{i}^{\prime}$ is an AE and can be given by
\begin{equation}
\boldsymbol{e}_{i}^{\prime}=\boldsymbol{e}_{i}+{\boldsymbol{r}}_{i}^{\prime}.
\end{equation}
To obtain such AE set $\mathcal{E}^{\prime}=\left\{\boldsymbol{e}^{\prime}_{i} \mid i=1, \ldots, t \right\}$, it is necessary to add perturbations in the direction where the current loss value increases the most, so as to make the largest change on the model prediction \cite{1555162}. That's to say, 
\begin{equation}
{\boldsymbol{r}^{\prime}}=\underset{\boldsymbol{r},\|\boldsymbol{r}\| \leq \epsilon}{\operatorname{argmax}}\{\mathcal{L}(\mathcal{E}^{\prime}, \mathcal{A},\boldsymbol{\Theta}^{\prime})\},
\end{equation}
where $\epsilon$ is a hyperparameter to control the norm of perturbations and $\boldsymbol{r}^{\prime}$ stands for the concatenated vector of $\boldsymbol{r}^{\prime}_{i}$ for all $i$. $\boldsymbol{\Theta}^{\prime}$ represents the current model parameters, enabling backpropagation \cite{rumelhart1986learning} not be used at the stage of AE generation.

Generally, it is infeasible to obtain a closed form for ${\boldsymbol{r}^{\prime}}$ since exact maximization with respect to $\boldsymbol{r}$ is intractable for DNNs. Fortunately, Goodfellow et al. \cite{1555162} introduce an efficient and fast gradient approximation  method by linearizing $\mathcal{L}(\mathcal{E}, \mathcal{A},\boldsymbol{\Theta}^{\prime})$ around $\mathcal{E}$ to provide a non-iterative solution for computing ${\boldsymbol{r}^{\prime}}$ with $L_{2}$ norm constraint:
\begin{equation}
{\boldsymbol{r}^{\prime}}=\epsilon \frac{\boldsymbol{g}}{\|\boldsymbol{g}\|_{2}}, \ \mathrm{where}\ \boldsymbol{g}=\nabla_{\mathcal{E}} \mathcal{L}(\mathcal{E}, \mathcal{A}, \boldsymbol{\Theta}).
\end{equation}
With the approximated adversarial perturbation ${\boldsymbol{r}^{\prime}}$, AE set  $\mathcal{E}^{\prime}=\left\{ \boldsymbol{e}_{i}^{\prime} \mid i=1, \ldots, t \right\}$  could be constructed. We jointly use both the original example set $\mathcal{E}$ and AE set $\mathcal{E}^{\prime}$ to update $\boldsymbol{\Theta}$ by minimizing the total loss as:
\begin{equation}
\boldsymbol{\Theta}\gets \mathcal{L}(\mathcal{E}, \mathcal{A}, \boldsymbol{\Theta})+\beta\mathcal{L}(\mathcal{E}^{\prime}, \mathcal{A}, \boldsymbol{\Theta}),
\end{equation}
\begin{equation}
\mathcal{L}\left(\mathcal{E}^{\prime}, \mathcal{A}, \boldsymbol{\Theta}\right)=\frac{1}{|\mathcal{E}^{\prime}|-1} \sum_{t} l\left(\mathcal{F}\left(\mathbf{e}_{t-1}^{\prime}; \boldsymbol{\Theta}\right), a_{t}\right),
\end{equation}
where $\beta$ is a hyperparameter that controls the relative importance of the normal training and AT. For better illustration, we summarize the entire training scheme of ATKT in \textbf{Algorithm 1}.

%\begin{figure}[t]
%	\centering
%	\includegraphics[width=0.85\linewidth]{Fig/Fig3/Fig3.pdf}
%	\caption{Schematic illustration of total loss generation procedure by jointly using normal training and AT.}
%	%	\Description{A woman and a girl in white dresses sit in an open car.}
%\end{figure}

\begin{algorithm}[htb] 
	\caption{$:$ ATKT Training.} 
	\label{alg:Framwork} 
	\begin{algorithmic}[1] %这个1 表示每一行都显示数字
		\REQUIRE ~~\\ %算法的输入参数：Input
		Student interaction dataset $\mathcal{X}$.
		\ENSURE ~~\\ %算法的输出：Output
		Trained model $\mathcal{F}(;\boldsymbol{\Theta})$ with parameters $\boldsymbol{\Theta}$;
		\STATE Initialize $\boldsymbol{\Theta}$; 
		\label{ code:fram:extract }%对此行的标记，方便在文中引用算法的某个步骤
		\label{code:fram:trainbase}
		\WHILE {early stopping not reaches}
		\STATE Construct normal example set $\mathcal{E}$ using \emph{Interaction Projection} introduced in section 3.2.1;
		\STATE Calucate $\mathcal{L}(\mathcal{E}, \mathcal{A},\boldsymbol{\Theta})$ using equation (10);
		\STATE Construct AE set $\mathcal{E}^{\prime}$ using equation (11)-(13);
		\STATE Calucate $\mathcal{L}(\mathcal{E}^{\prime}, \mathcal{A},\boldsymbol{\Theta})$ using equation (15);
		\STATE $\boldsymbol{\Theta}\gets \{\mathcal{L}(\mathcal{E}, \mathcal{A}, \boldsymbol{\Theta})+ \beta\mathcal{L}(\mathcal{E}^{\prime}, \mathcal{A}, \boldsymbol{\Theta})\}$;
		\ENDWHILE
		%		\FOR{$j=1$ to $M$}
		%		\STATE $\mathbf{C}_{j} \leftarrow \mathcal{F}_{HED} (\mathbf{T}_{j})$;
		%		\STATE Concatenate $\mathbf{T}_{j}$ and $\mathbf{C}_{j}$ along the channel dimension to form a multi-channel input $\mathbf{M}_{j}$;
		%		\STATE $\mathbf{G}^{'}_{j} \leftarrow \mathcal{F}_{PANet} (\mathbf{M}_{j})$;
		%		\ENDFOR
		\RETURN $\boldsymbol{\Theta}$ %算法的返回值
	\end{algorithmic}
\end{algorithm}

\begin{table}[h]
	\caption{Statistics of the four benchmark datasets.}
	\begin{tabular}{llll}
		\hline Dataset & Students & KSs & Responses \\
		\hline Statics2011 & 333 & 1,223 & 189,297 \\
		ASSISTments2009 & 4,151 & 110 & 325,637 \\
		ASSISTments2015 & 19,840 & 100 & 683,801 \\
		ASSISTments2017 & 1,709 & 102 & 942,816 \\
		\hline
	\end{tabular}
\end{table}

\begin{table*}
	\caption{Prediction performance of student's future responses in terms of various methods. The best two results are highlighted in {\color{red} red} and {\color{blue} blue}, respectively.}
	\begin{tabular}{llllllll}
		\hline Dataset & BKT+ \cite{yudelson2013individualized} & DKT \cite{NIPS2015_bac9162b} & DKT+ \cite{yeung2018addressing}& DKVMN \cite{zhang2017dynamic}& SAKT \cite{c4e83c715e3a43c596941347a55c91d7}& AKT \cite{ghosh2020context}& ATKT (Our)\\
		\hline 
		Statics2011 & $\sim 0.75$ & $0.8233 \pm 0.0039$ & {\color{blue}0.8301 $\pm$ 0.0039} & $0.8195 \pm 0.0041$ & $0.8029 \pm 0.0032$ & $0.8265 \pm 0.0049$ & {\color{red}0.8325 $\pm$ 0.0043}\\
		ASSISTments2009 & $\sim 0.69$ & {\color{blue}0.8170 $\pm$ 0.0043} & $0.8024 \pm 0.0045$ & $0.8093 \pm 0.0044$ & $0.7520 \pm 0.0040$ & $0.8169 \pm 0.0045$ & {\color{red}0.8244 $\pm$ 0.0032} \\
		ASSISTments2015 & & $0.7310 \pm 0.0018$ & $0.7313 \pm 0.0018$ & $0.7276 \pm 0.0017$ & $0.7212 \pm 0.0020$ & {\color{blue}0.7828 $\pm$ 0.0019} & {\color{red}0.8045 $\pm$ 0.0097}\\
		ASSISTments2017 & & $0.7263 \pm 0.0054$ & $0.7124 \pm 0.0041$ & $0.7073 \pm 0.0044$ & $0.6569 \pm 0.0027$ & {\color{blue}0.7282 $\pm$ 0.0037} & {\color{red}0.7297 $\pm$ 0.0051} \\
		\hline
	\end{tabular}
\end{table*}

\section{Experiment}
%In this section, we implement extensive experiments to verify the effectiveness of the proposed method. We first introduce the benchmark datasets, baseline methods and evaluation metric used in this work. We then report the performacne of student furure response prediction 

%on four real-world benchmarks
\subsection{Experimental Setting}
\subsubsection{Benchmark Datasets}
We use four widely-used benchmark datasets to evaluate the prediction performance of student's future responses. These datasets are introduced below. 

\textbf{$\bullet$ Statics2011\footnote{https://pslcdatashop.web.cmu.edu/DatasetInfo?datasetId=507}}: This dataset is provided by a college-level engineering statics course containing 189,927 responses with 1223 KSs from 333 students. It has the most KSs out of the four datasets.

\textbf{$\bullet$ ASSISTments2009\footnote{https://sites.google.com/site/assistmentsdata/home and \\ https://sites.google.com/view/assistmentsdatamining/\label{web}}}: This dataset is gathered from the ASSISTments online tutoring platform, which contains 525,637 responses with 110 KSs from 4151 students. 

\textbf{$\bullet$ ASSISTments2015}\textsuperscript{\ref {web}}: This dataset comes from the same platform as ASSISTments2009, but has a larger number of students and responses. Specifically, it includes 683, 801 responses with 100 KSs from 19,840 students. 

\textbf{$\bullet$ ASSISTments2017}\textsuperscript{\ref {web}}: This dataset is also from the same platform as ASSISTments2009. It has 942,816 responses with 102 KSs from 1709 students. Among the four datasets, this dataset contains the largest number of responses.

The detailed statistics of these datasets are listed in Table 1. We use the preprocessed version\footnote{https://github.com/arghosh/AKT} of these four datasets provided by \cite{1555162} to make a fair comparison. 

\subsubsection{Baseline Methods and Evaluation Metric}
To demonstrate the superiority of our ATKT, we compare our proposal with the following six state-of-the-art methods.

\textbf{$\bullet$ BKT+}
\cite{yudelson2013individualized}: This method is a BKT based method that takes student personalization into account. 

\textbf{$\bullet$ DKT}
\cite{NIPS2015_bac9162b}: This method is the first attempt to apply DNNs to KT, where a RNN model is employed to estimate student performance. 

\textbf{$\bullet$ DKT+}
\cite{yeung2018addressing}: This method is an improved version of DKT that uses prediction-consistent regularization technique to resolve the problems of input reconstruction failure and prediction inconsistency in DKT. 

\textbf{$\bullet$ DKVMN}
\cite{zhang2017dynamic}: This method is based on the memory-augmented neural networks where a static matrix is designed to store the KSs and a dynamic matrix is presented to store and update the mastery levels of corresponding KSs. 

%This method is based on the memory-augmented neural networks where the designed static matrix is to store the knowledge comcepts while the presented static matrix contribute to store and update the mastery levels of corresponding concepts. 

\textbf{$\bullet$ SAKT}
\cite{c4e83c715e3a43c596941347a55c91d7}: This method is a self-attention based method that assigns weights to the previous interactions for the future performance prediction. 

\textbf{$\bullet$ AKT} \cite{ghosh2020context}: This method is also an attention based method where the monotonic attention mechanism is proposed to capture the connections between student's current interaction and  previous interactions. 

Among these methods, BKT+ is the traditional machine learning based method while DKT, DKT+, DKVMN, SAKT and AKT are DNNs based methods. All of these methods form a solid baseline to measure the proposed ATKT comprehensively. Following DKT, SAKT, AKT, we use the Area Under Curve (AUC) metric to measure the prediction performance of each method. The value range of AUC is $[0, 1]$, where an AUC value of 0.5 indicates that the model performance is equal to a random guess. The larger the value, the better the performance of the method.

\subsubsection{Implementation Details}
Following \cite{ghosh2020context}, we perform standard 5-fold cross-validation on all datasets for our method. For each fold, the partition ratio of the training set, validation set and test set is set to 3:1:1. For a fair comparison, both the partitioned dataset and the prediction performance of other methods are taken from \cite{ghosh2020context}. Note that there are two variants of AKT in \cite{ghosh2020context}. The first variant uses KS information to model KT while the second one employs both KS and question information to accomplish KT. Since our main focus in this work is to explore the use of AT to enhance the generalization of the KT model, we only use KS information to model KT. Hence, we only compare with the first variant of AKT for fairness. 

The KS embedding dimension $d_{s}$ and response embedding dimension are empirically set to 256 and 96, respectively. We use a unidirectional single-layer LSTM with 80 hidden units to establish the attentive-LSTM model. For the proposed KHS attention module, $d_{w}$ is set to 80, and $d_{h}$ is equal to the number of hidden units in the LSTM, i.e., 80. Similar to \cite{ghosh2020context}, we truncate student interaction sequences longer than 500. For training, the batch size on all datasets is set to 24. We use the Adam optimizer \cite{DBLP:journals/corr/KingmaB14} with an initial learning rate of 0.001 and a 0.5 learning rate decay factor for 50 epochs. We perform the early stopping technique \cite{prechelt1998automatic} to finish training in advance when the loss on the validation set is no longer decreasing for 20 epochs. The maximum number of training epochs is set to 150. We use validation sets to tune the combination of the hyperparameters $\epsilon$ and $\beta$ (see section 4.3.1 for more details). We implement our experiments on the popular platform PyTorch \cite{NEURIPS2019_bdbca288} using 2 NVIDIA GeForce RTX 3090 GPUs.

\subsection{Comparison with State-of-the-Arts}
We show the prediction performance of various methods in Table 2 where the averages and standard deviations across five test folds are reported. It can be observed that the performance of all DNNs based KT methods are superior to that of BKT+, a representative traditional machine learning based method. Among the DNNs based KT methods, the performance of DKT and DKT+ is similar and SAKT illustrates the baseline. We can also see that the proposed ATKT consistently outperforms other methods on all datasets. Specifically, ATKT beats the close competitors DKT+ and DKT on the Statics2011 and ASSISTments2009 dataset, respectively. Both ATKT and AKT outperform other methods on the ASSISTments2015 dataset by a large margin while the proposed ATKT yields the best prediction performance. On the ASSISTments2017 dataset, both AKT and our method achieve the best two results but ATKT gives a better one. Those quantitative results demonstrate that our ATKT achieves a better prediction performance of student's future response.

\subsection{Ablation Studies}
In this section, we conduct thorough ablation studies to justify the effectiveness of the key components of our ATKT, including the implementation of AT and the proposed attensive-LSTM backbone for KT.

\begin{table}
	\caption{Average AUC scores on four validation sets with different combinations of $\epsilon$ and $\beta$. Note that $\beta=0$ represents no use of AT implementation. The best results are highlighted in bold.} 
	\begin{tabular}{cccccc}
		\hline \multirow{2}{*} {Statics2011} & \multicolumn{5}{c}{AUC}  \\ 
		\cline {2 - 6} & $\beta=0$ &$\beta=0.2$& $\beta=0.5$ & $\beta=1$ & $\beta=2$\\
		\hline 
		$\epsilon=1$ &0.8316&0.8318&0.8319&0.8323&0.8324\\
		$\epsilon=5$ &0.8316&0.8324&0.8322&0.8322&0.8317\\
		$\epsilon=10$ &0.8316&\textbf{0.8327}&0.8318&0.8309&0.8295\\
		$\epsilon=12$ &0.8316&\textbf{0.8327}&0.8315&0.8303&0.8282\\
		$\epsilon=15$ &0.8316&0.8324&0.8313&0.8291&0.8270\\
		\hline \hline \multirow{2}{*} {ASSISTments2009} & \multicolumn{5}{c}{AUC}  \\ 
		\cline {2 - 6} & $\beta=0$ &$\beta=0.2$& $\beta=0.5$ & $\beta=1$ & $\beta=2$\\
		\hline
		$\epsilon=1$ &0.8235&0.8233&0.8244&0.8243&0.8244\\
		$\epsilon=5$ &0.8235&0.8242&0.8249&0.8246&0.8245\\
		$\epsilon=10$ &0.8235&\textbf{0.8251}&0.8243&0.8244&0.8217\\
		$\epsilon=12$ &0.8235&0.8249&0.8245&0.8238&0.8196\\
		$\epsilon=15$ &0.8235&0.8248&0.8249&0.8227&0.8185\\
		\hline \hline \multirow{2}{*} {ASSISTments2015} & \multicolumn{5}{c}{AUC}  \\ 
		\cline {2 - 6} & $\beta=0$ &$\beta=0.2$& $\beta=0.5$ & $\beta=1$ & $\beta=2$\\
		\hline 
		$\epsilon=1$ &0.7792&0.7841&0.7842&0.7879&0.7888\\
		$\epsilon=5$ &0.7792&0.7863&0.7912&0.7966&0.7939\\
		$\epsilon=10$ &0.7792&0.7885&0.7924&0.8004&0.7975\\
		$\epsilon=12$ &0.7792&0.7906&0.7971&0.8001&0.7971\\
		$\epsilon=15$ &0.7792&0.7912&0.7994&\textbf{0.8053}&0.7995\\
		\hline \hline \multirow{2}{*} {ASSISTments2017} & \multicolumn{5}{c}{AUC}  \\ 
		\cline {2 - 6} & $\beta=0$ &$\beta=0.2$& $\beta=0.5$ & $\beta=1$ & $\beta=2$\\
		\hline 
		$\epsilon=1$ &0.7228&0.7236&0.7235&0.7241&0.7244\\
		$\epsilon=5$ &0.7228&0.7249&0.7265&0.7270&0.7282\\
		$\epsilon=10$ &0.7228&0.7262&0.7273&0.7293&0.7294\\
		$\epsilon=12$ &0.7228&0.7262&0.7281&\textbf{0.7297}&0.7291\\
		$\epsilon=15$ &0.7228&0.7267&0.7289&\textbf{0.7297}&0.7280\\
		\hline
	\end{tabular}
\end{table}

\begin{table}
	\caption{Average AUC scores on four test sets obtained with and without AT implementation, respectively.} 
	\begin{tabular}{ccc}
		\hline \multirow{2}{*} {Method}& \multicolumn{2}{c}{AUC}  \\ 
		\cline {2 - 3} & w/o AT & w/ AT\\
		\hline Statics2011 &0.8317&$\textbf{0.8325}$\\
		ASSISTments2009 &0.8230&$\textbf{0.8244}$\\
		ASSISTments2015 &0.7788&$\textbf{0.8045}$\\
		ASSISTments2017 &0.7224&$\textbf{0.7297}$\\
		\hline
	\end{tabular}
\end{table}

\begin{figure}[t]
	\centering
	\includegraphics[width=\linewidth]{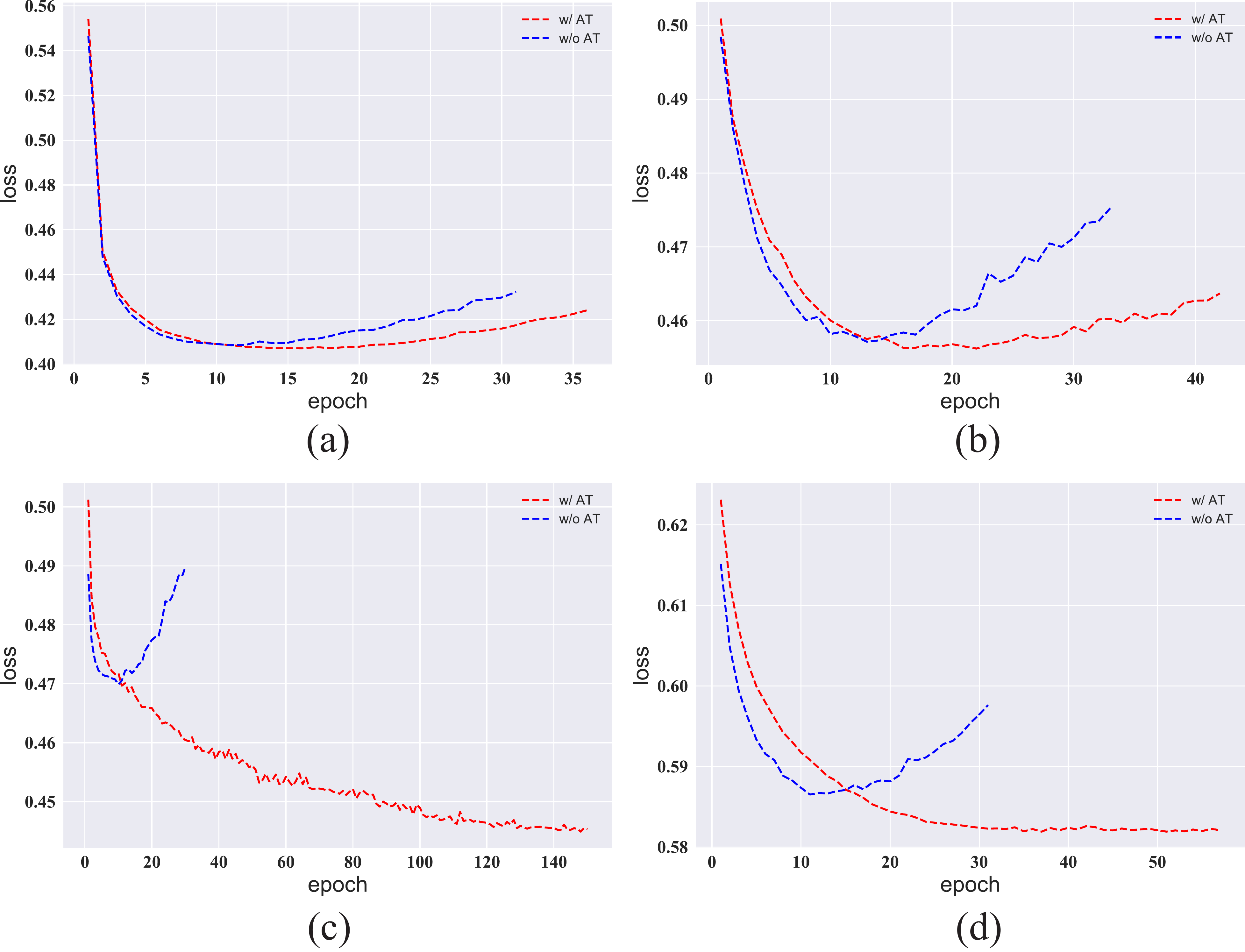}
	\caption{Validation loss curves on four benchmark datasets with and without AT implementation. (a)-(d): validation loss curve on the dataset of Statics2011, ASSISTments2009, ASSISTments2015 and ASSISTments2017, respectively.}
	%	\Description{A woman and a girl in white dresses sit in an open car.}
\end{figure}

\subsubsection{Effectiveness of the AT Implementation}
To implement AT, two hyperparameters should be noted. As described in section 3.3, the first one is $\epsilon$ that controls the norm constraint of perturbations and the second one is regularization weight $\beta$, which makes a tradeoff between normal training and AT. A larger $\epsilon$ indicates a stronger perturbation while a larger $\beta$ encourages KT framework to pay more focus on the AT loss. We set $\epsilon=\{1, 5, 10,12,15\}$ and $\beta=\{0, 0.2,0.5,1,2\}$ to search the ``\emph{optimal}'' hyperparameter combination on various validation sets. We first show the validation loss curves in Fig. 4, where we can see that the use of AT can efficiently alleviate KT framework risk of overfitting, especially on the large datasets such as ASSISTments2015 and ASSISTments2017. We then show the hyperparameter searched results in terms of AUC scores in Table 3 where $\beta=0$ means no use of AT implementation. 

It can be seen that for each $\epsilon$, the trend of the AUC score increases first and then decreases as the value of $\beta$ increases. This shows that the implementation of AT can efficiently improve the AUC scores; however, if the value of $\beta$ is too large, the AUC performance may degrade or even lower than the case of without AT implementation. For the small datasets such as Statics2011 and ASSISTments2009, the best AUC score would be achieved with a smaller $\beta$. This may be because the proposed KT backbone always has a greater risk of overfitting on small datasets. A larger $\beta$ would increase the influence of AT loss, the robustness of KT model may be damaged to a certain extent and its generalization on validation sets is also somewhat limited. As for the large datasets like ASSISTments2015 and ASSISTments2017, since our KT backbone has a less risk of overfitting on them, a larger $\beta$ may enhance the generalization of the KT backbone, leading to better AUC performance. With the searched ``\emph{optimal}'' hyperparameter combination, we also report the model's AUC performance on the test sets in Table 4, from which we can see that the AUC scores increase across various test sets by implementing AT. In a nutshell, the use of AT can enhance our KT backbone's generalization and result in better AUC performance.

\begin{figure*}[t]
	\centering
	\includegraphics[width=1.0\linewidth]{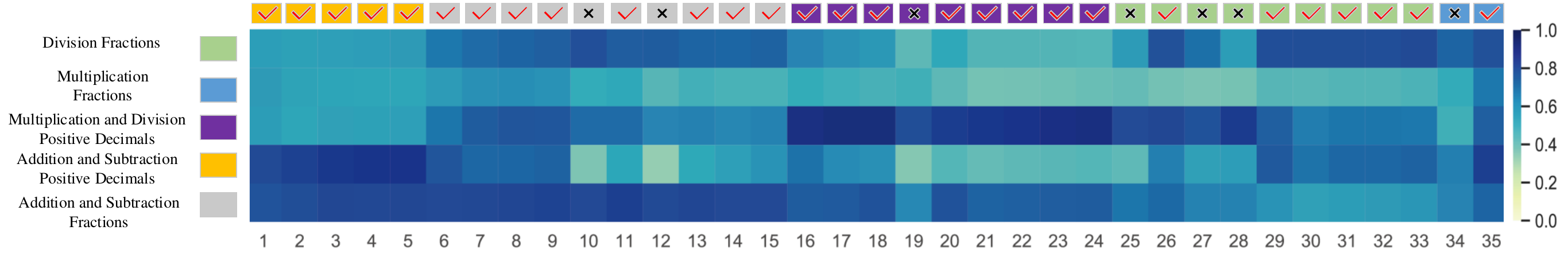}
	\caption{Visualization of the predicted student's mastery of 5 KSs over 35 steps during the exercising process using our ATKT. Top row records the student's performance on each exercise and each color denotes a distinct KS that the exercise contains. The heatmap in the middle is the student's KS mastery degree over steps where a deeper color represents a better mastery.}
	%	\Description{A woman and a girl in white dresses sit in an open car.}
\end{figure*}

\begin{figure}[t]
	\centering
	\includegraphics[width=0.8\linewidth]{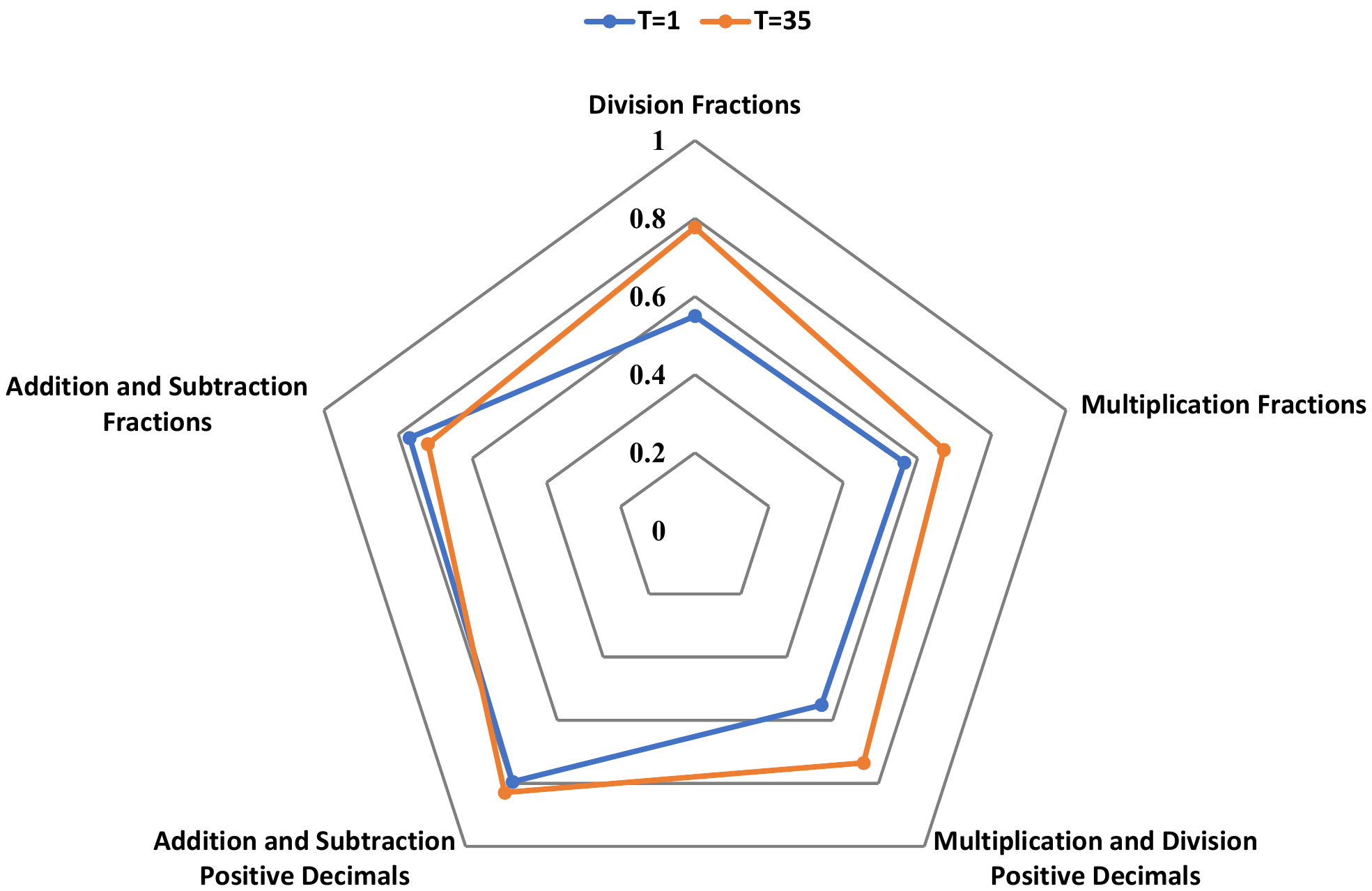}
	\caption{Comparison of the final (t=35) and initial (t=1) student's mastery of each KS.}
	%	\Description{A woman and a girl in white dresses sit in an open car.}
\end{figure}

\subsubsection{Effectiveness of the proposed Attensive-LSTM}
In our work, the KT backbone is the proposed  attentive-LSTM model, where the key is a customized KHS attention module that adaptively aggregates information from previous KHSs while simultaneously highlighting the importance of current KHS to make a more accurate prediction. To verify the effectiveness of this module, we train an attentive-LSTM model and a corresponding degraded one (a sole LSTM model) without KHS attention module to accomplish KT, respectively. The quantitative results are shown in Table 5. We can observe that equipped with our KHS attention module, the attentive-LSTM model efficiently improves the average AUC scores across all the datasets, demonstrating its superiority in boosting the performance of KT. 

\begin{table}
	\caption{Average AUC scores on four test sets obtained with and without the proposed KHS attention module.} 
	\begin{tabular}{ccc}
		\hline \multirow{2}{*} {Method}& \multicolumn{2}{c}{AUC}  \\ 
		\cline {2 - 3} & w/o KHS attention & w/ KHS attention\\
		\hline Statics2011 &0.8294&$\textbf{0.8317}$\\
		ASSISTments2009 &0.8167&$\textbf{0.8230}$\\
		ASSISTments2015 &0.7282&$\textbf{0.7788}$\\
		ASSISTments2017 &0.7222&$\textbf{0.7224}$\\
		\hline
	\end{tabular}
\end{table}

\subsection{Visualization}
In order to intuitively observe the KT performance of the proposed ATKT, we here visualize the predicted student's mastery of various KSs at each step during the exercising process. Due to the limited space, we only visualize a certain student's 5 KSs changing process over 35 exercises. The visualization result is exhibited in Fig. 5. Top row records the student's performance on each exercise and each color denotes a distinct KS that the exercise contains. The heatmap in the middle is the student's KS mastery degree over steps where a deeper color represents a better mastery. The heatmap in the middle is the student's KS mastery degree over steps where a deeper color represents a better mastery.

At the beginning, the student's KS mastery for each KS is set to 0 and will be updated through exercising. If he/she responds correctly to a specific exercise, the mastery of KSs will be improved accordingly, vice verse. For instance, when the student correctly answers the first five exercises with the same KS ``Addition and Subtraction Positive Decimals'', the mastery degree shown in the fourth row is increasing correspondingly. During the exercising, we can observe that the mastery degree of a certain KS would be affected by other ones. Take ``Multiplication Fractions'' as an example. The student does exercises with this KS only at 34 and 35 steps, but the predicted mastery degree of this KS changes gradually before 34 steps. This is mainly because there are potential connections between different KSs and thus  they may influence each other's mastery updates during the exercising. After exercising, we can obtain the student's final mastery of each KS. To have a direct comparison between the final KS mastery and the initial one, we show both of them in Fig. 6, where we can see that the student masters well in terms of the KSs ``Addition and Subtraction Positive Decimals'', ``Division Fractions'', ``Multiplication and Division Positive Decimals'' and ``Multiplication Fractions'', but has a relatively low mastery for the KS ``Addition and Subtraction Fractions''. With the visualization results, students can know how much they have mastered certain KSs and then carry out targeted exercises. 

\section{Conclusion}

In this paper, we explored the use of AT to model KT. First, We presented an efficient attentive-LSTM model as KT backbone to better fulfill KT. We then constructed the interaction embeddings as an input representation of KT to produce adversarial perturbations for the generation of AEs. Both the original clean inputs and AEs were jointly used to train our KT backbone, enhancing not only the robustness to the AEs but generalization for original ones. We conducted extensive experiments to evaluate our ATKT. Results revealed that ATKT is superior to currently state-of-the-art methods in terms of the AUC metric, demonstrating the effectiveness of AT in KT modelling. We also visualized the predicted student's mastery degree for each KS during the exercising process using our ATKT. We believe our observation of modelling KT in an AT manner would encourage future research.

\begin{acks}
The authors would like to thank the anonymous reviewers for their valuable comments and helpful suggestions. This work is supported by the National Natural Science Foundation of China under Grant No.: 62071014.
\end{acks}

\balance 
\bibliographystyle{ACM-Reference-Format}
\bibliography{sample-base}
\balance 
\appendix

\end{document}